\documentclass[prd,showpacs,twocolumn,
amsmath,amssymb,superscriptaddress,floatfix,nofootinbib]{revtex4-1}
\usepackage[utf8]{inputenc}
\usepackage[sort&compress]{natbib}
\usepackage[normalem]{ulem}
\usepackage{lipsum} 
\usepackage{bm}

\usepackage{times}
\usepackage{amssymb,amsbsy,amsmath,amsfonts}
\usepackage{graphicx}
\usepackage{float}
\usepackage{color}
\usepackage{morefloats}
\usepackage{rotating}
\usepackage{srcltx}
\usepackage{slashed}
\usepackage{subfigure}
\usepackage{multirow}
\usepackage{verbatim}
\usepackage{tabularx}
\usepackage{adjustbox}
\usepackage[dvipsnames]{xcolor}
\usepackage{soul}
\setstcolor{red}
\usepackage{nicefrac}
\usepackage{array}
\usepackage{makecell}
\usepackage{multirow}

\usepackage{hyperref}
\hypersetup{
	colorlinks	=true,
	urlcolor	=blue,
	linkcolor	=blue,
	citecolor	=blue,
	pdftitle	={},
	pdfauthor	={Zhi-Wei Liu, Jun-Xu Lu, Ming-Zhu Liu, Li-Sheng Geng},
	pdfsubject	={Deuteron-Deuteron Interaction and Correlation Function}
	}

\makeatletter

\newcommand{\Rmnum}[1]{\expandafter\@slowromancap\romannumeral #1@}
\makeatother

\begin{document}

\title{
Deuteron-Deuteron Interaction and Correlation Function
}

\author{Duo-Lun Ge}
\affiliation{School of Physics, Beihang University, Beijing 102206, China}

\author{Zhi-Wei Liu}
\email[Corresponding author: ]{liuzhw@buaa.edu.cn}
\affiliation{School of Physics, Beihang University, Beijing 102206, China}

\author{Jun-Xu Lu}
\affiliation{School of Physics, Beihang University, Beijing 102206, China}

\author{Li-Sheng Geng}
\email[Corresponding author: ]{lisheng.geng@buaa.edu.cn}
\affiliation{School of Physics, Beihang University, Beijing 102206, China}
\affiliation{Sino-French Carbon Neutrality Research Center, \'Ecole Centrale de P\'ekin/School of General Engineering, Beihang University, Beijing 100191, China}
\affiliation{Peng Huanwu Collaborative Center for Research and Education, Beihang University, Beijing 100191, China}
\affiliation{Beijing Key Laboratory of Advanced Nuclear Materials and Physics, Beihang University, Beijing 102206, China }
\affiliation{Southern Center for Nuclear-Science Theory (SCNT), Institute of Modern Physics, Chinese Academy of Sciences, Huizhou 516000, China}

\begin{abstract}
The interaction between deuterons ($d$-$d$) is pivotal for understanding the characteristics of certain light nuclei from the perspective of the deuteron cluster and achieving a precise reproduction of $d$-$d$ fusion cross sections. In this work, we construct a new set of elastic $d$-$d$ interactions by fitting the phase shifts using potentials parameterized in a Woods-Saxon shape. Then, the correlation functions are calculated with the obtained potential and compared with the recent measurements by the STAR collaboration. We find that the $d$-$d$ phase shifts and the correlation functions are internally consistent, confirming that correlation functions can provide cross-check for the $d$-$d$ interaction. In addition, both the $^1S_0$ bound state and the repulsive $^5S_2$ interaction contribute to the observed suppression in the measured correlation function. Moreover, we demonstrate that the $P$-wave contribution of the correlation functions cannot be neglected, especially in determining the source size.

\end{abstract}


\maketitle

\section{Introduction}
Unlike traditional scattering experiments, which often faced challenges in producing high-quality, short-lived particle beams, the femtoscopic technique can serve as a valuable supplement to constrain the strong interactions~\cite{Fabbietti:2020bfg, Liu:2024uxn}. More recently, a large variety of hadron-hadron interactions have been extensively studied by the ALICE and STAR collaborations, such as $p$-$K^\pm$~\cite{ALICE:2019gcn}, $p$-$\phi$~\cite{ALICE:2021cpv}, $p$-$\Xi^-$~\cite{ALICE:2019hdt}, $p$-$\Omega^-$~\cite{ALICE:2020mfd}, $\Lambda$-$\Lambda$~\cite{STAR:2014dcy}, and $\bar{p}$-$\bar{p}$~\cite{STAR:2015kha}.  Meanwhile, the experimental achievement has also triggered a large number of theoretical studies~\cite{Ohnishi:2016elb, Morita:2016auo, Haidenbauer:2018jvl, Kamiya:2019uiw, Kamiya:2022thy, Liu:2022nec, Liu:2023uly, Liu:2023wfo, Liu:2024nac, Geng:2025ruq, Vidana:2023olz, Molina:2023oeu, Khemchandani:2023xup, Torres-Rincon:2024znb, Yan:2024aap, Li:2024tvo, Liu:2025eqw}.

Upon the introduction of femtoscopy into particle and nuclear physics, a significant number of studies~\cite{Chitwood:1985zz,Boal:1986zz,Pochodzalla:1987zz,Chen:1987zz,Gong:1993zz,Chen:1987rsm,Gelderloos:1995zza,Elmaani:1994zz} concentrated on light nuclei, involving deuteron, triton, $\alpha$ particle, etc. For a review, one can see Ref.~\cite{Boal:1990yh}. However, the experimental data at that time lacked the precision to draw definitive conclusions. Recently, there has been a surge in highly accurate experimental measurements~\cite{Wang:2021mrv,Mi:2022zig,STAR:2024zvj}, reigniting interests and debates regarding the correlations involving light nuclei~\cite{Haidenbauer:2020uew, Ogata:2021mbo, ALICE:2023bny, Wang:2023ygv, Qiao:2024wha, Etminan:2024uvc, Torres-Rincon:2024znb}. These precise correlations not only provide a cross-check for the strong interactions obtained from scattering experiments, but also offer spatio-temporal information on the emission source.

The deuteron-deuteron ($d$-$d$) interaction is crucial in accurately reproducing the properties of certain nuclei within the cluster model~\cite{shojaei2015deuteron}. Although $\alpha$ clusters are most frequently utilized for this purpose, there have also been suggestions and explorations using deuteron and triton clusters in light $p$-shell nuclei and at the surface of closed shell core nuclei~\cite{Sakuta:2014qia,Taniguchi:2019awl,Bellini:2020cbj}. The $d$-$d$ interaction is also essential in accurately reproducing the $d$-$d$ fusion cross-section~\cite{Li_2008}. Moreover, the $d$-$d$ elastic scattering may reveal the contribution of the $I=0$ configuration to the observed discrepancies between the scattering data and the microscopic calculations ~\cite{Deltuva:2015fta}. However, it is precisely because the other four-body channels, such as $^3$H-$p$ and $^3$He-$n$, are already open at the $d$-$d$ threshold that the study of the $d$-$d$ interaction becomes very complicated. Given these considerations, some discrepancies arise in theoretical calculations of the $d$-$d$ scattering lengths~\footnote{A positive scattering length $a$ denotes either a repulsive interaction or the presence of a bound state, while a negative one signifies an attractive potential. However, this relationship breaks down for strongly attractive potentials yielding multiple bound states.}. For instance, the calculations employing a cluster reduction of the Yakubovsky components equations~\cite{filikhin2000investigation,filikhin2000microscopic} yield $10.2-0.2i$ fm for $^1S_0$ and $7.5$ fm for $^5S_2$, while the two-body cluster EFT~\cite{farzin2023low} results are $4.3170$ fm for $^1S_0$ and $6.1428$ fm for $^5S_2$. Additionally, another approach using the Faddeev-Yakubovskii chain-of-partition momentum-space equations finds the $^5S_2$ scattering length to be $7.8\pm0.3$ fm~\cite{Carew:2021jai}. These inconsistencies underscore the need for a more thorough and systematic study of the $d$-$d$ interaction.

Despite the inconsistencies mentioned above, with the accumulation of $d$-$d$ scattering data and the progress in phase shift analysis over the past few decades, a more accurate description of the $d$-$d$ potential becomes possible.
For example, a comprehensive partial-wave decomposition of the $d$-$d$ phase shifts has been performed by Hofmann and Hale~\cite{Hofmann:1996jv}, which includes $S$, $P$, and $D$ waves. At the same time, the femtoscopic technique has made impressive progress. In recent years, there has been a notable increase in highly accurate experimental measurements of $d$-$d$ correlation functions~\cite{Wang:2021mrv,Mi:2022zig,STAR:2024zvj}, achieving greater precision compared to the earlier works~\cite{Chitwood:1985zz,Chen:1987zz,Gong:1993zz}.

Given the recent advancements in measuring the $d$-$d$ correlation functions, it appears timely to reassess the currently available $d$-$d$ potentials reported in Ref.~\cite{Chitwood:1985zz}, which were originally derived using phase shift data~\cite{hale1984few,chwieroth1972study}. New findings on phase shifts~\cite{Hofmann:1996jv}, along with precise experimental results for the $d$-$d$ correlation functions~\cite{Wang:2021mrv,Mi:2022zig,STAR:2024zvj,Qiao:2024wha}, now allow for a more accurate analysis of the $d$-$d$ interaction. Therefore, this work aims to determine whether the $d$-$d$ correlation function can serve as a cross-check for the $d$-$d$ interaction obtained from scattering data and to provide additional information on the deuteron emission source.

The paper is organized as follows: Sect.~\ref{Sec:II} presents the theoretical formalism for extracting the $d$-$d$ interaction through phase shift analysis, along with the fitted results for the $d$-$d$ interaction. In Sect.~\ref{Sec:III}, we describe the theoretical framework for correlation functions and perform a comparative analysis of the recent data from the STAR experiment~\cite{STAR:2024zvj}. Finally, Sect.~\ref{Sec:IV} summarizes our key findings and provides an outlook for future research.

\section{Deuteron-Deuteron Interaction}\label{Sec:II}
We determine the elastic $d$-$d$ interaction by fitting it to the corresponding phase shifts, neglecting couplings to the $p$-$3N$ system. We first solve the single-channel Schr\"odinger equation for two-body scattering in configuration space:
\begin{align}\label{Eq:SE}
    -\frac{\hbar^2}{2\mu}\nabla^2\psi(\mathbf{r})+V(r)\psi(\mathbf{r})=E\psi(\mathbf{r}),
\end{align}
where $\mu=m_1m_2/(m_1+m_2)$ is the reduced mass of the system, and $E=\hbar^2k^2/2\mu$ is the total energy. Here, the potential $V(r)$ is assumed to be of the widely used Woods-Saxon form
\begin{align}\label{Eq:V_Woods-Saxon}
    V(r)=\frac{V_0}{1+e^{\frac{r-R}{a}}},
\end{align}
where $V_0$, $R$, and $a$ are parameters that determine the potential's depth, range, and steepness, respectively. The total wave function can be separated into a radial part and an angular part after partial-wave decomposition, i.e., $\psi(r,\theta,\varphi)=\sum_{l=0}^{l_{\rm max}}R_l(r)Y^l_m(\theta,\varphi)=\sum_{l=0}^{l_{\rm max}}u_l(r)Y^l_m(\theta,\varphi)/r$. The $l_{\rm max}$ is determined by the convergence of the sum over the orbital angular momentum. For each partial wave, the radial Schr\"odinger equation becomes
\begin{align}\label{Eq:SER}
    -\frac{\hbar^2}{2\mu}\frac{\mathrm{d}^2u_l(r)}{\mathrm{d}r^2}+\left[\frac{\hbar^2}{2\mu}\frac{l(l+1)}{r^2}+V(r)\right]u_l(r)=Eu_l(r).
\end{align}

Once the $d$-$d$ wave function is obtained, one can calculate the corresponding phase shifts. The free radial wave function is proportional to the spherical Bessel function $j_l(pr)$, which behaves as $\sin(pr-l\pi/2)$. For a short-range interaction, the radial wave function $u_l(r)$ in the asymptotic region is shifted by an amount $\delta_l(p)$, known as the phase shifts for a specific partial wave. By comparing the asymptotic wave function to the initial wave function, one can accurately determine the phase shifts $\delta_l(p)$ for that particular partial wave. With the Coulomb interaction, the overall process for determining the phase shifts remains unchanged except for replacing the free asymptotic wave function with the regular Coulomb wave function.

\begin{table}[htbp]
\caption{Parameters of the Woods-Saxon potential determined by fitting the phase shifts in Fig.~\ref{Fig:dd_PS_Fit} for various partial waves permitted by quantum statistics. The positive $V_0$ indicates a repulsive interaction, and the negative value represents an attractive potential. The last two columns represent the binding energy and the scattering length, respectively.}
\centering
\begin{tabular}{ccccccc}
\hline
\hline
\textbf{Data Set} & \textbf{Channel} & $\mathbf{V_0}$ \textbf{(MeV)} & \textbf{R (fm)} & \textbf{a (fm)} & \textbf{B (MeV)} & $\mathbf{a_0}$ \textbf{(fm)} \\ \hline
\multirow{5}{*}{\makecell[c]{\textbf{RM}}}
                  & $\mathbf{^1S_0}$ & $-56.38$ & $8.23$ & $0.45$ & $52.89$ & $-0.26$ \\
                  & $\mathbf{^5S_2}$ & $29.71$ & $1.10$ & $1.37$ & -- & $3.96$ \\
                  & $\mathbf{^3P_0}$ & $1.09$ & $8.78$ & $1.19$ & -- & -- \\
                  & $\mathbf{^3P_1}$ & $1.00$ & $9.01$ & $1.35$ & -- & -- \\
                  & $\mathbf{^3P_2}$ & $6.15$ & $1.05$ & $1.83$ & -- & -- \\
                  \hline
\multirow{5}{*}{\makecell[c]{\textbf{RG}}}
                  & $\mathbf{^1S_0}$ & $-59.31$ & $2.97$ & $0.10$ & $42.09$ & $4.95$ \\
                  & $\mathbf{^5S_2}$ & $35.74$ & $2.23$ & $0.92$ & -- & $3.55$ \\
                  & $\mathbf{^3P_0}$ & $59.57$ & $1.61$ & $1.24$ & -- & -- \\
                  & $\mathbf{^3P_1}$ & $82.93$ & $1.86$ & $1.20$ & -- & -- \\
                  & $\mathbf{^3P_2}$ & $79.86$ & $1.04$ & $1.37$ & -- & -- \\
                  \hline \hline
\end{tabular}
\label{Tab:interaction_potentials}
\end{table}

\begin{figure*}[htbp]
  \centering
  \includegraphics[width=0.98\textwidth]{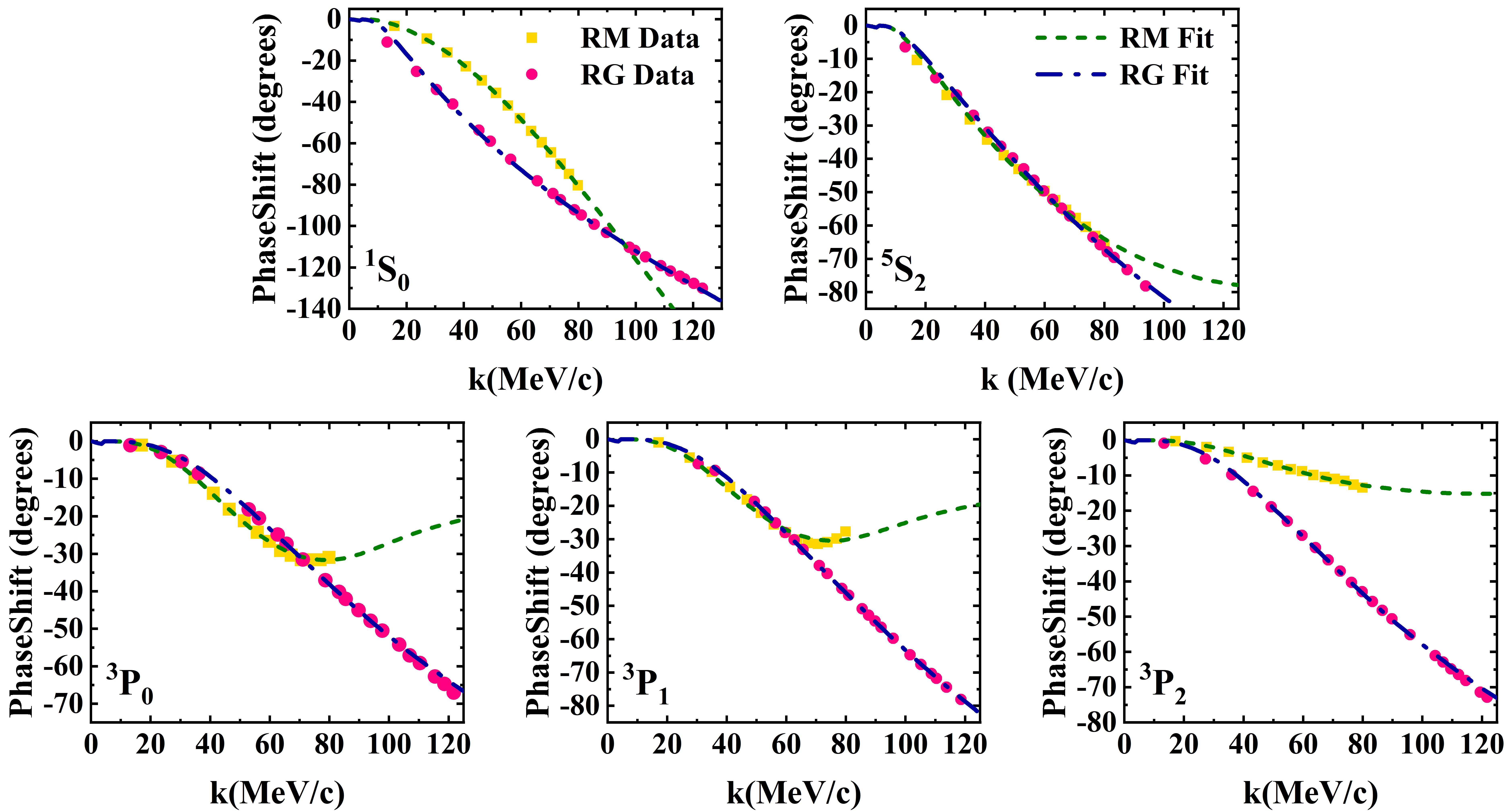}
  \caption{The $d$-$d$ phase shifts as a function of the relative momentum $k$ for different partial waves $^1S_0$, $^5S_2$, and $^3P_{0,1,2}$. The dots and squares are taken from Ref.~\cite{Hofmann:1996jv}, while the dashed and dash-dotted lines show the fitted results of the present work. See the main text for more details.}\label{Fig:dd_PS_Fit}
\end{figure*}

Fig.~\ref{Fig:dd_PS_Fit} shows the fitted results for each partial wave. The colored squares and dots represent two sets of data, denoted as RM and RG~\cite{Hofmann:1996jv}, respectively. The RM data were obtained using the R-matrix analysis, which has only the data in the $^4$He system as input. In contrast, the RG data were obtained using the Resonating Group Model (RGM), which employed the r-space version of the Bonn potential —a realistic nucleon-nucleon potential with no adjustable parameters. Notably, the differential cross-sections of all 4N scattering processes were well reproduced with the RGM. The squares and dots in Fig.~\ref{Fig:dd_PS_Fit} are taken from Ref.~\cite{Hofmann:1996jv} after transforming the center-of-mass energy $E_{\rm c.m.}$ to the relative momentum $k$. Nearly every point is perfectly fitted by the dashed and dash-dotted lines. The extracted potential parameters for these two sets of phase shifts are given in Table~\ref{Tab:interaction_potentials}. We obtain multiple best-fitting results during the fitting process, but their properties differ. As discussed in Ref.~\cite{PhysRevC.103.024318}, the $d$-$d$ potential is strongly attractive in the $^1S_0$ channel and repulsive in $^5S_2$. We have selected a set of fitted results applicable to the $^5S_2$ phase shifts in both datasets (RM and RG), with a $\chi^2$ value comparable to the minimum. For the $^1S_0$ channel, we stick with the minimization of the $\chi^2$ and find that the interaction for $^1S_0$ is strongly attractive, which is consistent with the conclusion of Ref.~\cite{PhysRevC.103.024318}. For $P$ waves, given the absence of bound states in the 4N system in this energy region, we only consider the repulsive results. For each partial wave, the binding energy is calculated by solving the Schr\"odinger equation for bound states, with results presented alongside the potential parameters and the $S$-wave scattering lengths in Table~\ref{Tab:interaction_potentials}.

\section{Deuteron-Deuteron Correlation Functions}\label{Sec:III}
Now, we turn to the $d$-$d$ correlation function, which is computed using the Koonin-Pratt equation~\cite{Koonin:1977fh,Pratt:1990zq}:
\begin{align}\label{Eq:K-P_function}
    C(k) = \int S(\mathbf{r})|\psi_k(\mathbf{r})|^2\mathrm{d}^3r,
\end{align}
where $k=|\mathbf{p}_1-\mathbf{p}_2|/2$ is the reduced relative momentum of the $d$-$d$ pair in the center-of-mass frame, $\mathbf{r}$ is the relative distance between the $d$-$d$ pair, $S(\mathbf{r})$ is the so-called source function, and $\psi_k(\mathbf{r})$ is the scattering wave function. In this work, we choose a normalized Gaussian distribution for the source function, which only depends on $r$
\begin{align}\label{Eq:Gaussian_source}
    S_G(r)=\frac{1}{(4\pi r_0^2)^{\frac{3}{2}}}\mathrm{exp}\,(-\frac{r^2}{4r_0^2}),
\end{align}
where $r_0$ is the size parameter of the source. The observed correlation function, which has been averaged over spin, is described by
\begin{align}\label{Eq:CATS_CF}
    C_{\rm tot}(k)=\sum_{S,L,J}c_{S,L,J}\cdot C_{S,L,J}(k),
\end{align}
where each $C_{S,L,J}(k)$ is evaluated through the Koonin-Pratt equation, $c_{S,L,J}=(2S+1)/((2s_1+1)(2s_2+1))\cdot(2J+1)/((2L+1)(2S+1))$ is the weight of each $\{S,L,J\}$ channel~\cite{Mihaylov:2018rva}, and  $s_1$, $s_2$ denote the spins of the two constituent particles, respectively. Since we are dealing with a system of identical bosons, quantum statistics dictates that only certain $\{S,L,J\}$ channels are allowed. Specifically, under the constraint that the total wave function must be symmetric, the exchange symmetry of the spatial and spin wave functions must be the same. Given that the spin symmetry obeys $(-1)^S$ for identical bosons, we only consider the partial waves satisfying $L+S =$ even. Considering both $S$ and $P$ waves, we obtain $^1S_0$, $^5S_2$, $^3P_0$, $^3P_1$, and $^3P_2$. For each allowed partial wave, the symmetrization and anti-symmetrization of the spatial wave function are given by
\begin{align}
    \Psi_k^{\rm sym}(\mathbf{r}) =& \frac{1}{\sqrt{2}}\left[\psi_k(\mathbf{r})+\psi_{k}(\mathbf{-r})\right],\\
    \Psi_k^{\rm anti-sym}(\mathbf{r}) =& \frac{1}{\sqrt{2}}\left[\psi_k(\mathbf{r})-\psi_{k}(\mathbf{-r})\right],
\end{align}
where $\psi_k(\mathbf{r})$ is the solution to Eq.~\ref{Eq:SE}. In practice, one must replace the wave function in Eq.~\ref{Eq:K-P_function} with the symmetrized or anti-symmetrized wave function $\Psi_k(\mathbf{r})$ depending on the spatial symmetry of the partial wave. It is worth mentioning that for the infinite-range Coulomb potential $V_c(r)=Q_1Q_2/r$, where $Q_1$ and $Q_2$ represent the dimensionless charges of the particle pair, the scattering wave function asymptotically behaves as the regular Coulomb wave function and will not converge to zero. Therefore, the integration over the entire space in the Koonin-Pratt equation (Eq.~\ref{Eq:K-P_function}) must be handled carefully. We have checked that the numerical integral is convergent in this work.

\begin{figure*}[htbp]
  \centering
  \includegraphics[width=0.98\textwidth]{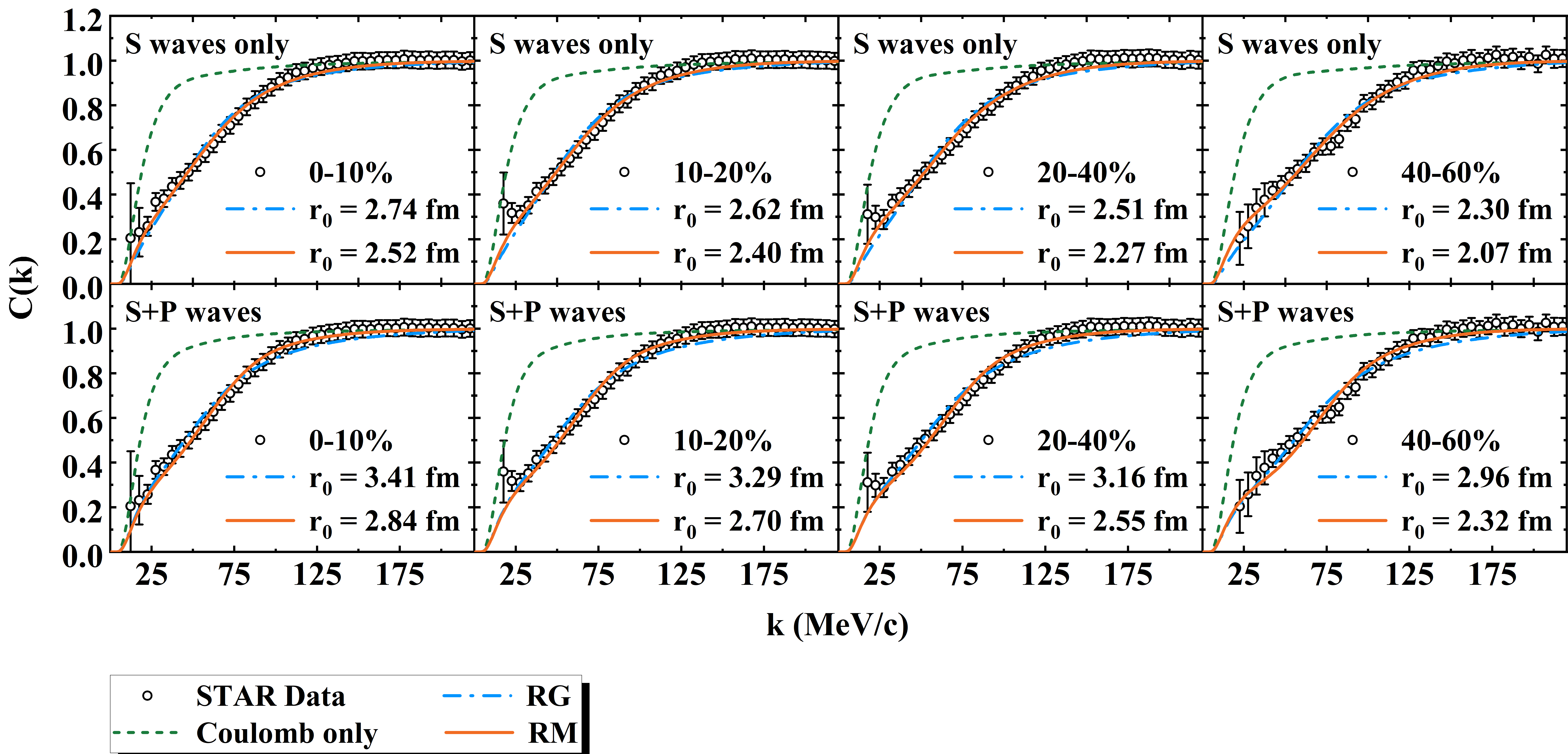}
  \caption{The $d$-$d$ correlation functions as a function of the relative momentum $k$ for different centralities. The dots are taken from Ref.~\cite{STAR:2024zvj}, and the error bars are determined by adding the systematic and statistical uncertainties in quadrature. The solid and dash-dotted lines show the fitted results using the RM and RG strong interactions, respectively. The dashed lines indicate the results obtained considering only the Coulomb potential. The first row includes strong interactions only in the $S$-wave, the second row employs the $P$-wave strong interaction. The source radii for the dashed and dash-dotted lines are identical.}\label{Fig:dd_CF_Fit}
\end{figure*}

The $d$-$d$ correlation functions are shown in Fig.~\ref{Fig:dd_CF_Fit}, representing the fitted results considering strong interactions only in the two $S$ waves (the first row) and all five partial waves (the second row) for different centralities. More specifically, in the first row, the three $P$ waves are accounted for solely through their Coulomb interactions, while the second row includes the strong interactions for the three $P$ waves. The fit was performed using the obtained RM/RG strong interactions shown in Table~\ref{Tab:interaction_potentials}, with the Gaussian source radii $r_0$ serving as the only tunable parameter. The solid and dash-dotted lines illustrate the fitted results using the RM strong and RG strong interactions, respectively. In contrast, the dashed lines depict pure Coulomb correlation functions using the same source radii as the solid lines. The disparity between the dashed and solid/dash-dotted lines thus predominantly manifests the $d$-$d$ strong interactions. In both scenarios, the correlation functions incorporating the strong interaction exhibit suppression relative to those considering only the Coulomb interaction and quantum statistics. This suppression in the $d$-$d$ correlation functions can be mainly attributed to the presence of $d$-$d$ bound states in $^1S_0$ and the repulsive interaction in $^5S_2$ for both datasets. To verify this interpretation, we conduct tests by turning off both the Coulomb potential and quantum statistical effects. Under these conditions, the behavior of the two $^1S_0$ correlation functions mirrors that of bound states, consistent with the findings for a strongly attractive interaction reported in Ref.~\cite{Liu:2023uly}, where the source size dependence on various interactions has been studied in a square-well model. 

Moreover,under current experimental accuracy, both the RM and RG $S$-wave interactions already reproduce the experimental correlation functions well, incorporating $P$-wave strong interactions further reduces the $\chi^2$ values for both cases. It should be noted that the RM and RG interactions differ substantially in the $^1S_0$ scattering lengths (Table~\ref{Tab:interaction_potentials}), yet their spin-averaged scattering lengths $\bar{a}_0=\frac{1}{6}a_0^{^1S_0}+\frac{5}{6}a_0^{^5S_2}$ show only minor variations. This explains their similar descriptions of spin-averaged correlation functions, since the experimental data inherently weight each partial wave contribution statistically. While these data cannot determine specific partial-wave interactions, they provide a cross-check for the overall spin-averaged interaction. Notably, the RM interaction excellently reproduces both phase shifts and correlation functions, which demonstrates their internal consistency and validates the observation in Ref.~\cite{Hofmann:1996jv}.

It is noteworthy that a similar methodology was previously attempted in Ref.~\cite{Chitwood:1985zz}, where a Woods-Saxon type $d$-$d$ interaction was extracted through phase shifts fitting using data from Ref.~\cite{hale1984few,chwieroth1972study}. However, significant distinctions emerge in our current work. First, phase shift analysis and correlation function measurement techniques have undergone significant improvements over the past four decades. For instance, our analysis is built upon the comprehensive framework of Ref.~\cite{Hofmann:1996jv}, which adopted the refined RGM calculations~\cite{Ferreira1987ModelsAM}, and recent high-precision $d$-$d$ correlation data from the STAR collaboration~\cite{STAR:2024zvj}. Second, Ref.~\cite{Chitwood:1985zz} attributes the suppression in the $d$-$d$ correlation functions solely to the fitted repulsive $d$-$d$ strong interactions. We propose a fundamentally different physical interpretation. Specifically, we suggest that the observed suppression may also originate from the $d$-$d$ bound states in the $^1S_0$ channel. Therefore, the observed suppression is the collective result influenced by the bound state in $^1S_0$ and the repulsive $^5S_2$ interaction. This distinction is critical since repulsive potentials and bound-state effects can produce similar suppression patterns relative to pure Coulomb correlation functions.

\begin{figure*}[htbp]
  \centering
  \includegraphics[width=0.62\textwidth]{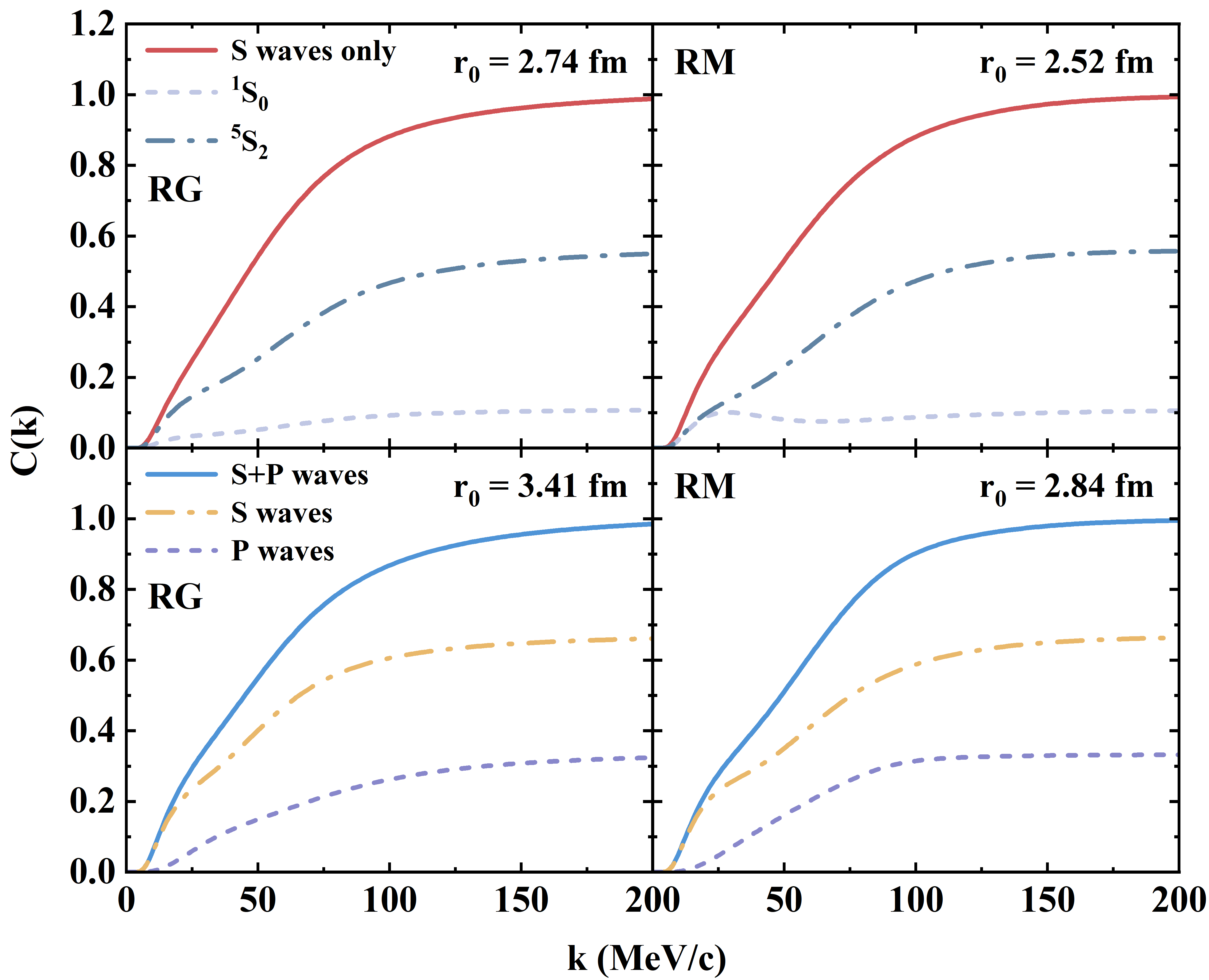}
  \caption{The $d$-$d$ correlation functions as a function of the relative momentum $k$ for 0-10\% centrality. The first row highlights contributions from $^1S_0$ and $^5S_2$, while the second row contrasts contributions from $S$ and $P$ waves. The left column illustrates scenarios based on the fitted RG strong interaction, and the right column shows those for the RM strong interaction.}\label{Fig:dd_CF_Fit_PWC}
\end{figure*}

Next, we present the contributions of various partial waves in Fig.~\ref{Fig:dd_CF_Fit_PWC}. The first row presents the outcomes based solely on $S$ waves, whereas the second row illustrates the results considering both $S$ and $P$ waves. In each panel, the solid line represents the total correlation function, and the dashed and dash-dotted lines show the contribution of various partial waves. In the first row, the $^1S_0$ and $^5S_2$ channels dominate with a combined weight of $2/3$, while the remaining $1/3$ originates from higher partial waves ($P$ waves, $F$ waves, etc., up to convergence) in the $S=1$ channel. Crucially, these higher waves are attributed solely to the Coulomb interactions, as discussed below Fig.~\ref{Fig:dd_CF_Fit}. As for the second row, the overall correlation function is simply the sum of the $S$-wave and $P$-wave contributions. 
For large relative momentum $k$, the correlation function for each partial wave is approximately the weight, i.e., $C_{S,L,J}(k) \xrightarrow{\text{large}\, k} c_{S,L,J}\times 1$. Consequently, the $S$ waves dominate, with the weight of $2/3$, especially the $^5S_2$ partial wave, which carries the weight of $5/9$, being predominant among the $S$ waves. The three $P$ waves comprise $1/3$ of the total weight, so their contribution should not be neglected.

\begin{figure}[htbp]
  \centering
  \includegraphics[width=0.48\textwidth]{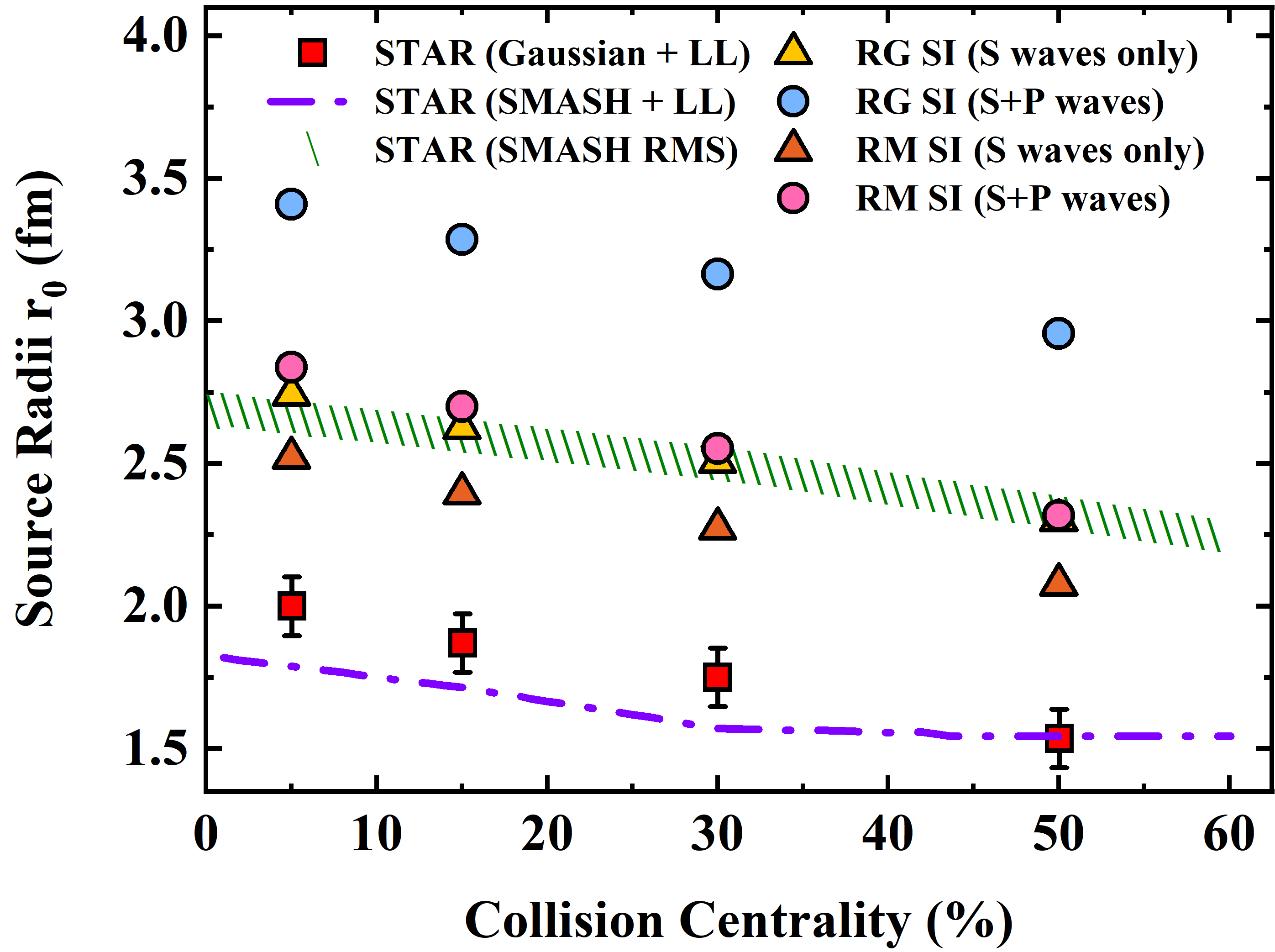}
  \caption{Collision centrality dependence of the Gaussian source radii. The triangles and the circles result from the strong interaction, including only two $S$ waves and all five partial waves. The red squares represent the results obtained by directly fitting the experimental data using only the LL model. All data points represent the center value of each bin, corresponding to the ranges 0-10\%, 10-20\%, 20-40\%, and 40-60\%, respectively. The dashed purple lines represent the outcomes of the SMASH model and the LL model simultaneously, and the green diagonal lines indicate the root mean square values of the source radii directly extracted from the SMASH model.
  }\label{Fig:dd_SR_Fit}
\end{figure}

In Fig.~\ref{Fig:dd_SR_Fit}, we present the fitted results of source radii compared with the recent STAR data. The triangles represent the results fitted with the $S$-wave strong interactions, while the circles were fitted with the interactions that include both $S$ and $P$ waves. The red squares represent the results obtained by fitting the data from different collision centralities using the Gaussian source plus the LL model~\cite{Lednicky:1981su,PhysRevC.101.015201}. Note that the bins for all five sets of data points (triangles, circles, and squares) are consistent, specifically 0-10\%, 10-20\%, 20-40\%, and 40-60\% (corresponding to Fig.~\ref{Fig:dd_CF_Fit}). Meanwhile, the purple dash-dotted line indicates the outcome of the SMASH model (a newly developed hadronic transport model)~\cite{PhysRevC.94.054905} plus the LL model. The SMASH model exclusively handles the phase space configuration of particles, whereas the LL model is responsible for implementing quantum statistics and final-state interactions. The green diagonal lines depict the root mean square values directly extracted from the SMASH model. It is clear that all results consistently show a decreasing trend as the collision centrality increases, which can be attributed to the reduction in the number of participating partons as centrality increases. We also discovered that including the three $P$ waves increases the Gaussian source radii. On the other hand, the slight difference in the source radii between the two $S$-wave interactions reflects their minor phase shift variations. When the $P$-wave interactions are included, the considerable difference in source radii originates primarily from the significant discrepancies in the $P$-wave phase shifts, particularly in the high-momentum region for the $^3P_0$ and $^3P_1$ channels, and across all momenta for the $^3P_2$ channel. Additionally, it is noteworthy that source radii extracted using the RG $S$-wave strong interactions show consistency with the green diagonal lines derived from the SMASH model (RMS). Furthermore, all our fitted source radii agree with the SMASH model (RMS) values compared to those directly obtained from the LL model fits. 

\section{Summary and outlook}\label{Sec:IV}
In this work, we constructed the deuteron-deuteron interactions with the Woods-Saxon form by fitting the phase shifts, including both $S$ and $P$ waves. The $^1S_0$ binding effect and the repulsive $^5S_2$ interaction are responsible for the observed suppression in the correlation functions. The internal consistency between phase shifts and correlation functions emerges from successfully reproducing $d$-$d$ phase shifts and correlation functions. Compared to the RG strong interaction, the correlation functions calculated with the RM strong interactions show better agreement with the experimental results. In addition, the partial-wave decomposition of the correlation function indicates that the contribution from the three $P$ waves cannot be overlooked, especially in determining the source size.

We employed a simple Gaussian distribution to characterize the emission source function in our calculations. Deuterons were treated as point-like objects without explicitly accounting for their composite nature and large size.  Refs.~\cite{Mrowczynski:2021bzy, Bazak:2020wjn,Mrowczynski:2020ugu}, on the other hand, treated the deuteron-deuteron system as a four-body system and found a source function of similar form but with a different source radius. We note that incorporating heavy-ion collision transport models~\cite{Wang:2024yke} or data-driven neural networks~\cite{Wang:2024bpl} as alternatives for the source function holds promise for further refining our description of experimental data. In addition, we plan to perform a detailed study involving the coupled-channel effects among proton-triton, deuteron-deuteron, and neutron-$^3$He systems.

\emph{Acknowledgments.} This work is partly supported by the National Natural Science Foundation of China under Grant No. 12435007. Zhi-Wei Liu acknowledges support from the National Natural Science Foundation of China under Grant No.12405133, No.12347180, China Postdoctoral Science Foundation under Grant No.2023M740189, and the Postdoctoral Fellowship Program of CPSF under Grant No.GZC20233381.

\bibliography{dd}
\clearpage

\end{document}